\documentclass{article}
\usepackage{spconf,amsmath,graphicx,hyperref,algorithm,
algorithmicx,tabularx,booktabs,multirow,makecell,subfigure}

\title{WiFiSim: Simulating WiFi Probe Requests via AOSP Analysis\\and Device Behavior Modeling}
\name{Lifei Hao, Yue Cheng, Min Wang, Bing Jia*\thanks{*Corresponding Author. Email: jiabing@imu.edu.cn} and Baoqi Huang}
\address{College of Computer Science, Inner Mongolia University\\
Engineering Research Center of Ecological Big Data, Ministry of Education}
\begin{document}
%
\maketitle

\begin{abstract}
WiFi probe request (PR) frames encode fine-grained device interactions and serve as a critical basis for mobility and crowd analytics. However, pervasive MAC address randomization and the scarcity of labeled datasets hinder progress in PR-based studies. We introduce \emph{WiFiSim}, a simulation framework that reconstructs PR generation through Android Open Source Project (AOSP) protocol analysis and finite-state device behavior modeling. WiFiSim identifies the key determinants of PR structure and timing while capturing realistic user-driven state transitions. Experiments show that WiFiSim achieves less than $5\%$ deviation from real measurements in both distributional and temporal dynamics, scales to large-scale dataset synthesis, and enables reliable evaluation of downstream applications. Source code and sample datasets are publicly released to foster reproducible research.
\end{abstract}

\begin{keywords}
WiFi Probe Request, MAC Randomization, AOSP, Behavior Modeling
\end{keywords}

\section{Introduction}
The ubiquity of smart devices and the evolution of WiFi standards have positioned WiFi probe request (PR) analysis as an enabler for smart cities, public safety, and business intelligence~\cite{hong2018crowdprobe}. A PR frame, broadcast during active scanning, carries attributes such as MAC address, received signal strength (RSS), and capability information. These features support device identification, crowd estimation, and mobility inference~\cite{10285068, yang2024privacy}. However, modern devices widely adopt MAC randomization to enhance privacy, rendering address-based identification ineffective and undermining device counting and trajectory tracking~\cite{fenske2021three}.  

Crucially, randomization does not fully eliminate privacy risks. Prior studies demonstrate that devices may still be re-identified by exploiting temporal patterns, RSS distributions, and protocol artifacts~\cite{matte2016defeating, bravenec2022exploration}, as well as information element (IE) fingerprints~\cite{tan2021efficient, he2022self} and machine learning–based classifiers~\cite{baccichet2024mac, 10225447}, yet all these methods critically depend on high-quality datasets, which remain limited. Existing collections often lack ground-truth labels, offer restricted scenario diversity, overlook behavioral state transitions, and are insufficient in scale for robust validation~\cite{zhao2023efficient, pintor2022dataset}. This underscores the need for labeled, realistic, and scalable PR datasets, which in turn raises three challenges: reconstructing realism PR generation and randomization; modeling user-driven scanning behavior with multi-state transitions; and enabling controllable, large-scale, multi-scenario synthesis.

To address these challenges, we present \emph{WiFiSim}, a framework for synthetic PR dataset generation. WiFiSim analyzes the Android Open Source Project (AOSP) protocol stack to extract the parameters and logic of PR construction, then adopts a finite-state machine (FSM) to model realistic user behaviors. By integrating analysis insights with behavior modeling, WiFiSim generates PR sequences with authentic temporal dynamics. Multi-device parallel simulation further enables scalable dataset synthesis with consistent ground truth. Compared to field data collection, WiFiSim provides complete labeling, high fidelity, and flexible scalability, establishing a reproducible foundation for downstream evaluation.  

Our main contributions are three-fold: 
(1) We propose the integration of AOSP protocol analysis and FSM-based modeling for high-fidelity PR simulation.  
(2) We validate the realism, scalability, and stability of WiFiSim datasets in evaluating downstream tasks such as device counting.  
(3) We release the source code and labeled sample datasets at \url{https://github.com/cyicz123/WiFiSim} to support reproducible and extensible WiFi mobility research.  

\section{Methodology}
We next describe the methodology of \emph{WiFiSim}, whose key idea is to integrate protocol stack analysis with device behavior modeling, thereby synthesizing PR traces with authentic structures, realistic dynamics, and complete ground truth.

\subsection{AOSP Source Code Analysis}
Android, as the dominant mobile platform, provides open-source implementations of the WiFi subsystem through the AOSP. By inspecting the source code, we identify the end-to-end process of PR generation, the configuration of critical parameters, and the mechanisms of MAC randomization.

\begin{figure}[htb]
    \centering
    \includegraphics[width=1\linewidth]{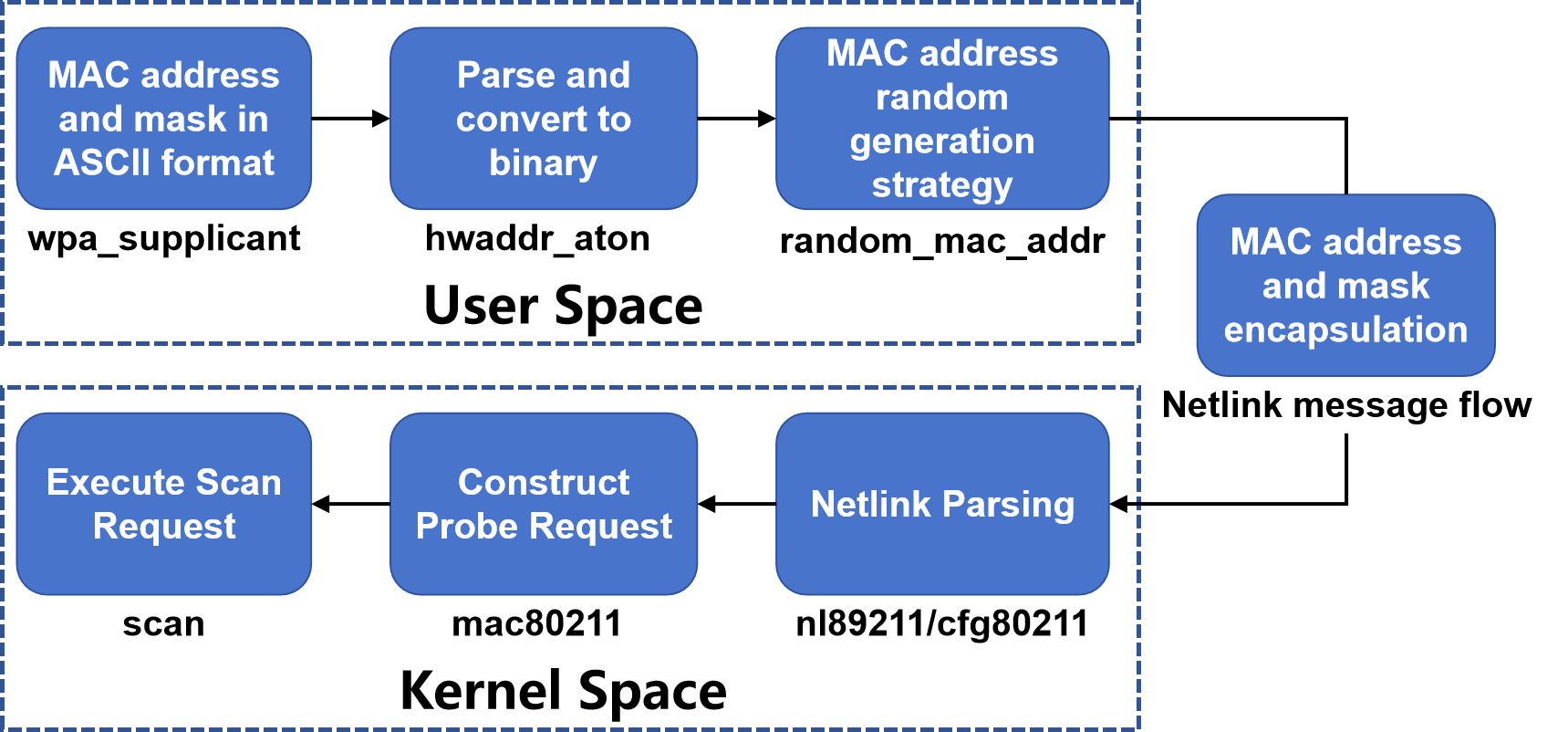}
    \caption{Pipeline of PR generation in AOSP WiFi stack.}
    \label{fig:pr_pipeline}
\end{figure}

The AOSP WiFi stack follows a layered design comprising the framework layer, the hardware abstraction layer (HAL), and the kernel space~\cite{repec:spr:prochp:978-3-031-48865-8_2}. The PR generation pipeline, illustrated in Fig.~\ref{fig:pr_pipeline}, can be separated into user-space and kernel-space operations. The \texttt{wpa\_supplicant} handles scan requests in user space, parses ASCII MAC strings into binary, and invokes \texttt{random\_mac\_addr()} to produce IEEE 802.11–compliant randomized addresses. These, combined with configuration data, are forwarded to the kernel via Netlink messages. Within kernel space, \texttt{cfg80211}/\texttt{nl80211} parse the requests, while \texttt{mac80211} constructs the PR frames prior to transmission.  

Overall, MAC randomization is executed in user space, whereas frame construction occurs in the kernel, with coordination ensured through the Netlink interface. Typical randomization strategies controlled by \texttt{wpa\_supplicant} include: (1) full randomization per scan, (2) OUI-preserving randomization, and (3) periodic randomization. All generated MAC must conform to IEEE 802.11 specifications~\cite{9363693}.

\begin{table}[htb]
\centering
\footnotesize
\caption{Representative WiFi scanning parameters captured and simulated by WiFiSim. (val.=value, prob.=probability)}
\label{tab:scan_params}
\renewcommand{\arraystretch}{1.1}
\begin{tabular}{ccc}
\hline
\textbf{Category} & \textbf{Field} & \textbf{Example/Description} \\
\hline
\multirow{3}{*}{\shortstack{Hardware\\Attributes}} 
  & Vendor           & Samsung \\
  & Device Model     & Galaxy S10 \\
  & OUI (in MAC)     & dc:ef:09 \\
\hline
\multirow{6}{*}{\shortstack{Protocol\\Capabilities}} 
  & Supported Rates (Mbps)       & 6:0.25 / ... / 18:0.25 \\
  & Ext. Supported Rates (Mbps)  & 24:0.5 / 36:0.5 \\
  & HT Capabilities (Hex)        & \texttt{98abcdef01234567} \\
  & VHT Capabilities (Hex)       & \texttt{ab12cd34ef567890} \\
  & Extended Capabilities (Hex)  & \texttt{12ab34cd56ef} \\
\hline
\multirow{3}{*}{\shortstack{Behavior\\Control}} 
  & Burst Length (val. : prob.)          & 1:0.3 / ... / 2:0.4 \\
  & MAC Policy (random type)            & OUI-preserving \\
\hline
\end{tabular}
\end{table}

For faithful emulation, \emph{WiFiSim} simulates three classes of parameters identified from AOSP analysis: \emph{hardware attributes} govern IEs embedded in PR frames, \emph{protocol capability} define the logical frame structure, and \emph{behavior control} determine temporal patterns of PR emission. Representative examples are shown in Table~\ref{tab:scan_params}. 

\subsection{User-Device Behavior Modeling}
Mobile devices frequently alternate between operational states, each exhibiting distinct scanning patterns. Following prior work~\cite{rusca2023privacy}, \emph{WiFiSim} employs an FSM to reproduce temporal heterogeneity across devices. Grounded on empirical usage statistics and Android policies, we define four principal states:
\vspace{-4pt}
\begin{itemize}
\setlength{\itemsep}{-4pt}
    \item \textbf{Shutdown:} Device is powered off or WiFi is disabled. No PRs are generated.
    \item \textbf{Screen-off:} Device is locked in low-power mode. Scanning occurs at long intervals with small bursts.
    \item \textbf{Screen-on:} Device is awake but idle. Scanning is moderately frequent, with intervals ranging from tens of seconds to minutes.
    \item \textbf{Activity:} Device is actively used. Scanning is highly frequent, with large bursts and short inter-burst delays.
\end{itemize}

\begin{figure}[htb]
    \centering
    \includegraphics[width=0.75\linewidth]{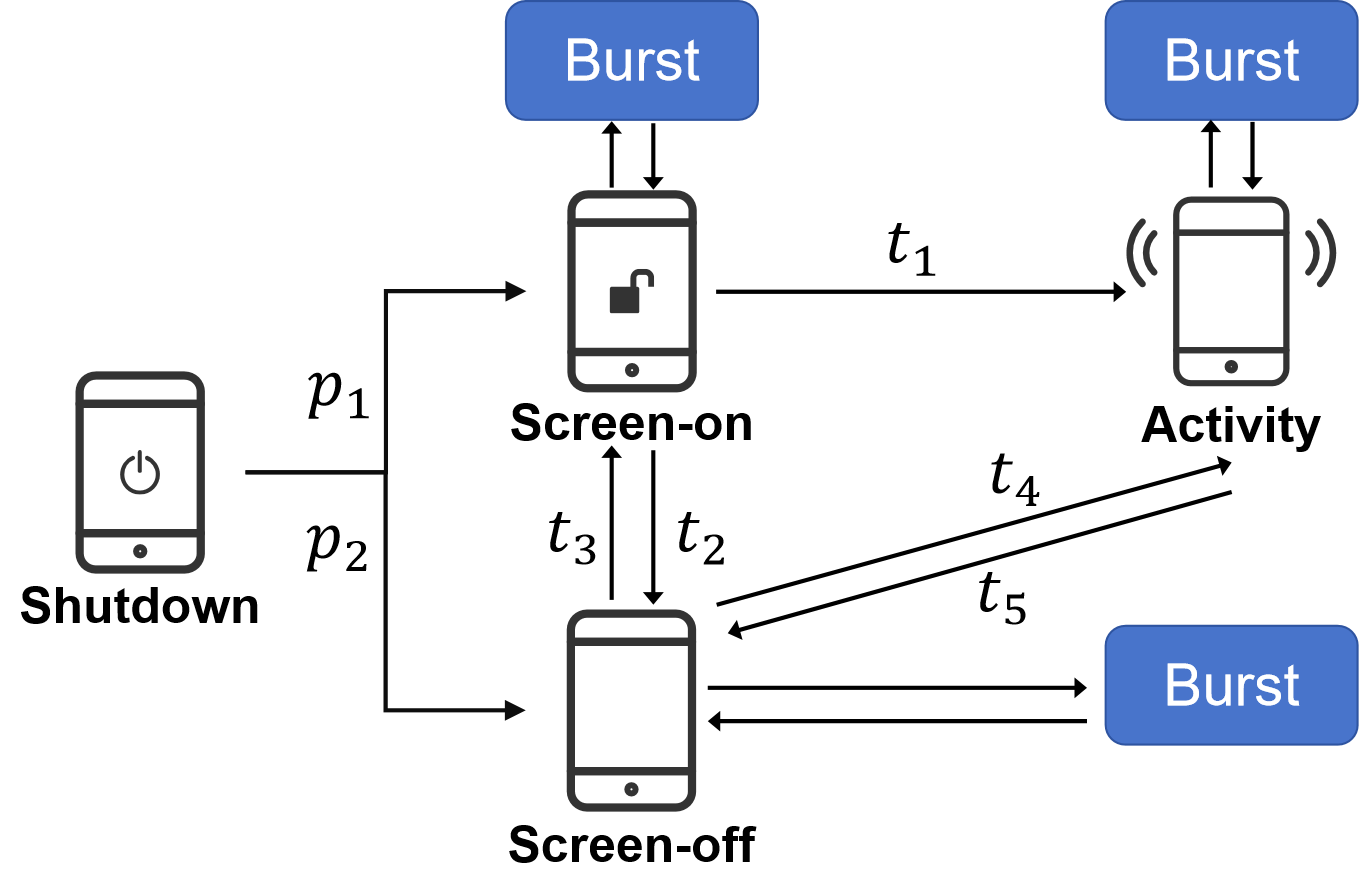}
    \caption{Device behavior modeling. ($p_i$ denotes transition probabilities and $t_i$ denotes temporal thresholds or events.)}
    \label{fig:fsm_transition}
\end{figure}

\begin{table*}[htb]
\centering
\footnotesize
\caption{Representative parameters for device behavior modeling. (val.=value, prob.=probability)}
\label{tab:behavior_params}
\renewcommand{\arraystretch}{1.1}
\begin{tabular}{cccc}
\hline
\textbf{Category} & \textbf{Field} & \textbf{Description} & \textbf{Example} \\
\hline
Device Identifier 
  & Device ID 
  & Unique index of a device   
  & Device 56 \\
\hline
\multirow{2}{*}{Device State} 
  & State 
  & \shortstack{Current mode (0: Shutdown, 1: Screen-off, 2: Screen-on, 3: Activity)} 
  & 1 \\
  & State Duration 
  & State duration distribution ($second$, val.:prob.) 
  & 20:0.7 / 30:0.3 \\
\hline
\multirow{3}{*}{\shortstack{Model-Related\\Scan Timing}} 
  & Intra-burst Interval 
  & Interval distribution within bursts ($second$, val.:prob.) 
  & 0.05:0.5 / 0.08:0.5 \\
  & Inter-burst Interval 
  & Distribution between bursts ($second$, val.:prob.) 
  & 3.0:0.6 / 4.0:0.4 \\
  & Jitter 
  & Random variation in intervals ($second$, val.:prob.)  
  & 0.05:0.5 / 0.1:0.5 \\
\hline
\end{tabular}
\end{table*}

To capture device and user diversity, probabilistic configurations are defined for each state, including durations, intra-/inter-burst intervals, and jitter. Configurations are exemplified in Table~\ref{tab:behavior_params}. State transitions follow the FSM in Fig.~\ref{fig:fsm_transition}, driven by user actions, system events, or temporal triggers. \emph{WiFiSim} augments this process with a Markovian transition model and time constraints to incorporate both randomness and diurnal periodicity (e.g., higher daytime activity).  

The FSM is executed through a discrete-time simulation (Algorithm~\ref{alg:fsm_sim}). At each step, the next state is sampled, corresponding scanning events are generated, and burst parameters are scheduled. Parameters are derived from empirical analysis to ensure fidelity, while the framework remains extensible for customized scenarios.

\begin{algorithm}[htb]
\footnotesize
\caption{FSM-driven Device Behavior Simulation}
\label{alg:fsm_sim}
\begin{algorithmic}[1]
\State \textbf{States:} \{Shutdown, Screen-off, Screen-on, Activity\}
\State \textbf{Transitions:}
    \State \quad Shutdown $\xrightarrow{p_1}$ Screen-on
    \State \quad Shutdown $\xrightarrow{p_2}$ Screen-off
    \State \quad Screen-on $\xrightarrow{t_1}$ Activity
    \State \quad Screen-on $\xrightarrow{t_2}$ Screen-off
    \State \quad Screen-off $\xrightarrow{t_3}$ Screen-on
    \State \quad Screen-off $\xrightarrow{t_4}$ Activity
    \State \quad Activity $\xrightarrow{t_5}$ Screen-off
\State \textbf{PR Trigger Parameters:}
    \State \quad Shutdown: \{scanInterval = 0s, WiFi = off\}
    \State \quad Screen-off: \{scanInterval = $600s\pm20\%$, WiFi = on\}
    \State \quad Screen-on: \{scanInterval = $120s\pm30\%$, WiFi = on\}
    \State \quad Activity: \{scanInterval = $30s\pm50\%$, WiFi = on\}
\end{algorithmic}
\end{algorithm}

\begin{table*}[htb]
\centering
\scriptsize
\caption{Sample simulated probe request frames displayed in Wireshark.}
\label{tab:PRexamples}
\begin{tabularx}{0.92\linewidth}{c c c c c c X}
\hline
\textbf{No.} & \textbf{Time (s)} & \textbf{Source} & \textbf{Destination} & \textbf{Prot.} & \textbf{Len.} & \makecell[c]{\textbf{Info}} \\
\hline
131993 & 638.2472 & OPPOdigital\_a3:e3:36             & Broadcast & 802.11 & 90  & Probe Req, SN=1576, RSS=-39dBm, SSID=Wildcard \\
131994 & 638.2499 & MS-NLB-PhysServer-18:34:56:78:9a  & Broadcast & 802.11 & 100 & Probe Req, SN=2486, RSS=-70dBm, SSID=Wildcard \\
131995 & 638.2751 & 8a:79:1c:43:15:da                 & Broadcast & 802.11 & 108 & Probe Req, SN=2723, RSS=-50dBm, SSID=Wildcard \\
131996 & 638.2889 & 62:d2:8e:0a:ae:ee                 & Broadcast & 802.11 & 122 & Probe Req, SN=1003, RSS=-67dBm, SSID="qnkBB9..." \\
131997 & 638.2942 & OnePlusElect\_6c:5a:40            & Broadcast & 802.11 & 90  & Probe Req, SN=3148, RSS=-40dBm, SSID=Wildcard \\
131998 & 638.3829 & 8a:09:3a:e8:61:f2                 & Broadcast & 802.11 & 150 & Probe Req, SN=1308, RSS=-31dBm, SSID="PhHmXR..." \\
131999 & 638.3838 & Lenovo\_cf:e1:24                  & Broadcast & 802.11 & 90  & Probe Req, SN=1088, RSS=-35dBm, SSID=Wildcard \\
\hline
\end{tabularx}
\end{table*}

\subsection{Scalable and Generalizable PR Dataset Construction}
\begin{figure}[htb]
    \centering
    \includegraphics[width=\linewidth]{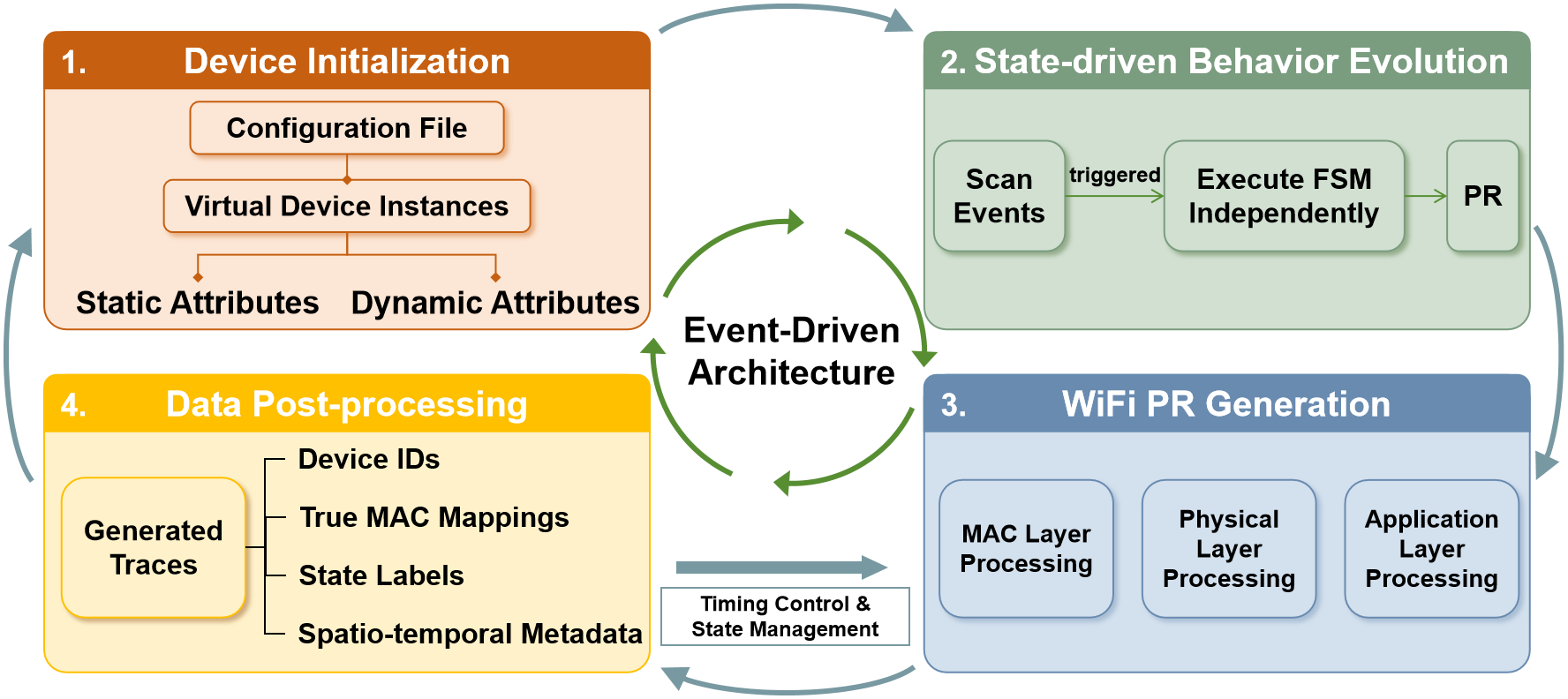}
    \caption{Workflow of \emph{WiFiSim} for PR dataset generation.}
    \label{fig:wifisim_framework}
\end{figure}

Building on the protocol and behavior models, \emph{WiFiSim} introduces a complete dataset construction framework that supports multi-device parallel emulation. The framework adopts an event-driven architecture to ensure fine-grained timing control and efficient state management, as depicted in Fig.~\ref{fig:wifisim_framework}. The workflow proceeds as follows:
\vspace{-4pt}
\begin{enumerate}
\setlength{\itemsep}{-5pt}
    \item \textbf{Device initialization:} Virtual device instances are instantiated from configuration files, each containing static attributes (e.g., vendor, OUI, protocol support) and dynamic attributes (e.g., randomization policy, scan intervals).
    \item \textbf{State-driven behavior:} Each device independently executes its FSM. Upon scan triggers, corresponding PR generation routines are activated.
    \item \textbf{PR generation:} The MAC layer applies address randomization, protocol parameters populate frame fields, and the physical layer simulates radio effects including path loss, multipath fading, and interference. The application layer sets scanning frequency.
    \item \textbf{Post-processing:} Generated traces are annotated with device IDs, state labels, true MAC mappings, and spatio-temporal metadata. Standardized datasets are then exported in PCAP format.
\end{enumerate}
\vspace{-4pt}

\begin{table*}[htb]
\centering
\footnotesize
\caption{Performance of device counting methods on WiFiSim-generated datasets.}
\label{tab:derand_practicality_wide}
\begin{tabular}{c|ccc|ccc|ccc}
\hline
\multirow{2}{*}{\textbf{Methods}} & 
\multicolumn{3}{c|}{\textbf{LMSS}} & 
\multicolumn{3}{c|}{\textbf{HMMS}} & 
\multicolumn{3}{c}{\textbf{HMLS}} \\
\cline{2-10}
& Acc.$\uparrow$ & MAE$\downarrow$ & MSE$\downarrow$ 
& Acc.$\uparrow$ & MAE$\downarrow$ & MSE$\downarrow$ 
& Acc.$\uparrow$ & MAE$\downarrow$ & MSE$\downarrow$ \\
\hline
IECluster~\cite{uras2020wifi} & 0.64 & 4.76 & 32.44 & 0.27 & 6.33 & 51.37 & 0.43 & 7.35 & 62.46 \\
IETime~\cite{bravenec2022exploration} & 0.77 & 1.37 & 2.89 & 0.77 & 1.08 & 2.12 & 0.75 & 1.43 & 3.02 \\
TimeRSS~\cite{yang2024privacy} & 0.81 & 1.11 & 2.57 & 0.85 & 0.91 & 1.31 & 0.79 & 1.19 & 2.73 \\
\hline
\end{tabular}
\end{table*}

\begin{figure*}[htb]
    \centering
    \subfigure[MRCA]{
        \includegraphics[width=0.242\linewidth]{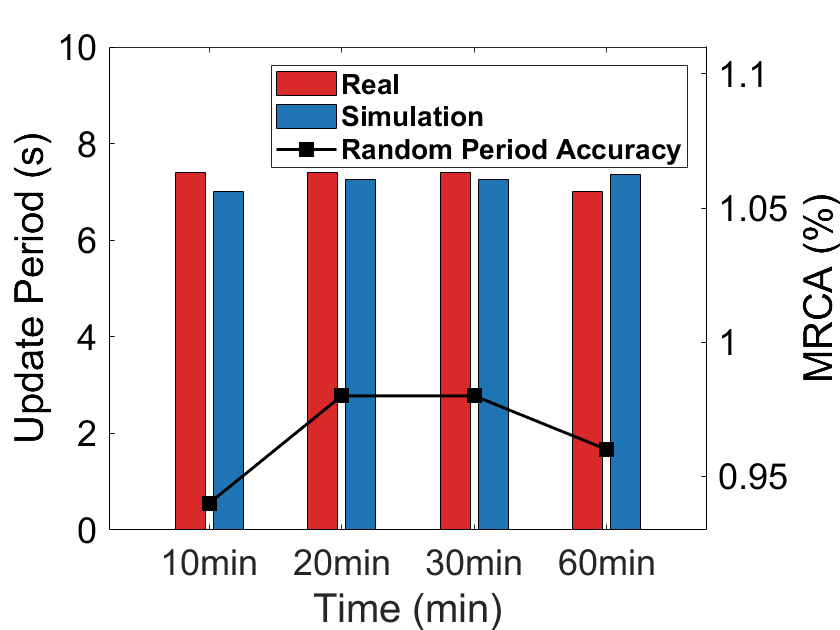}
    }
    \subfigure[MCR]{
        \includegraphics[width=0.242\linewidth]{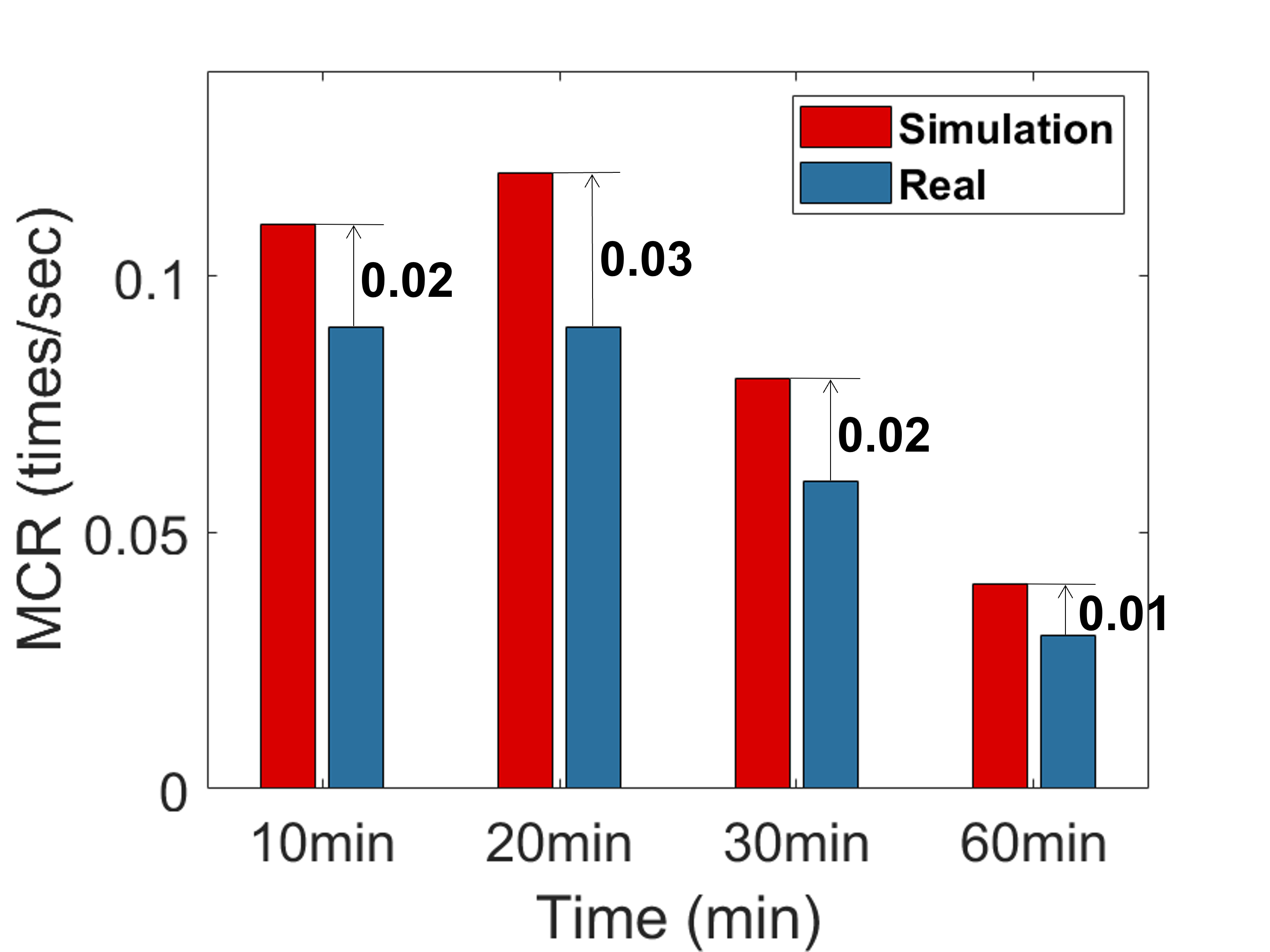}
    }
    \subfigure[NUMR]{
        \includegraphics[width=0.242\linewidth]{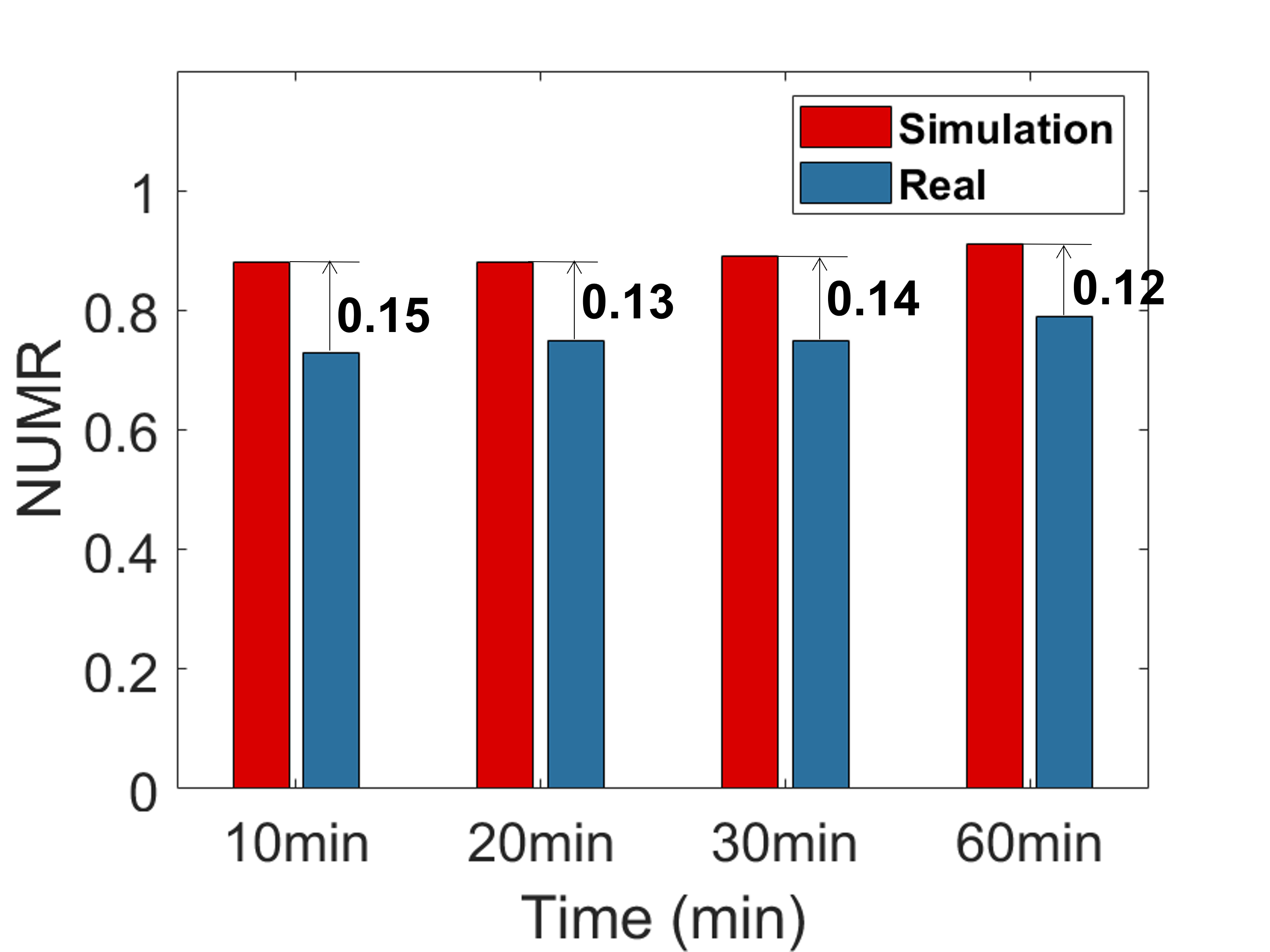}
    }
    \caption{Comparison of simulated and real traces under different observation windows.}
    \label{fig:realism_exp}
\end{figure*}

To enable large-scale evaluation, \emph{WiFiSim} supports flexible parameterization, scenario customization (e.g., number of devices, spatial density, mobility models), and multi-threaded execution. Table~\ref{tab:PRexamples} presents a subset of synthetic traces rendered in Wireshark~\cite{wireshark}. Compared with conventional field collection, \emph{WiFiSim} generates datasets of arbitrary scale with complete ground truth, substantially enhancing reproducibility for supervised learning and algorithm benchmarking.

\section{Experiments}
We evaluate \emph{WiFiSim} from two perspectives: (1) \emph{authenticity}, i.e., the similarity between simulated and real probe requests; and (2) \emph{practicality}, i.e., whether WiFiSim can serve as reliable input for downstream tasks such as device counting via MAC de-randomization.

\subsection{Validation of Data Authenticity}
WiFiSim-generated traces are compared with real PRs collected from a controlled setup. A Redmi Note 13 smartphone was placed in a semi-anechoic chamber, while an external WiFi adapter in monitor mode captured all 802.11 management frames using Wireshark with ground truth logging. To ensure comparability, the simulated device was configured with identical hardware OUI, protocol capabilities, MAC policies, and software parameters. Four observation windows were tested: 10, 20, 30, and 60 minutes, with state transitions between \emph{screen-off} and \emph{activity}.

Three metrics are designed for evaluation: (1) MAC Randomization Cycle Accuracy (MRCA) quantifies the alignment between simulated and expected randomization periods, reflecting cycle fidelity; (2) MAC Change Rate (MCR) measures the frequency of address changes per window, indicating randomization intensity; (3) Normalized Unique MAC Ratio (NUMR) denotes the ratio of total PRs to distinct MAC addresses, capturing address stability.

As shown in Fig.~\ref{fig:realism_exp}, MRCA remains above 95\% across all windows, confirming faithful reproduction of randomization cycles. MCR of simulated traces is slightly higher than observed traces, attributable to model simplifications at boundary conditions; however, this also highlights WiFiSim’s generalization capacity for emulating diverse vendor implementations. NUMR trends are consistent across all durations. Overall, deviations across the three metrics remain below 5\%, demonstrating the realism and stability of WiFiSim.

\subsection{Validation of Data Practicality}
We further assess the practicality of \emph{WiFiSim} for evaluating device counting methods via MAC de-randomization. This task associates PRs emitted under randomized MACs back to their originating devices, relying on temporal and protocol-level features with consistent ground truth.

Three representative baselines are reproduced: IECluster~\cite{uras2020wifi} (2020), an early method clustering randomized PRs directly from IEs; IETime~\cite{bravenec2022exploration} (2022), which augments IE features with temporal statistics and employs density-based clustering; TimeRSS~\cite{yang2024privacy} (2024), which leverages timestamps and RSS for unsupervised clustering. To provide broad coverage, we synthesize three datasets with varying scale and mobility: \emph{LMSS} (low-mobility, small-scale): 34 devices, 196 randomized MACs, 12 real MACs, duration 1\,h; \emph{HMMS} (high-mobility, mid-scale): 203 devices, 11,890 randomized MACs, 25 real MACs, duration 1\,h; \emph{HMLS} (low-mobility, large-scale): 247 devices, 1,392 randomized MACs, 38 real MACs, duration 1\,h. We adopt a 40\,s observation window and compute device counting accuracy (Acc.) as the proportion of estimates within $\pm5\%$ of ground truth. Mean absolute error (MAE) and mean squared error (MSE) are also reported, with lower values indicating higher accuracy and robustness.  

Results in Table~\ref{tab:derand_practicality_wide} indicate strong alignment between performance on WiFiSim datasets and algorithmic design. Two key findings emerge: (1) \emph{WiFiSim} preserves temporal and protocol-layer properties sufficient to reproduce relative performance trends across baselines.  
(2) Dataset characteristics affect methods differently—IE-only clustering degrades sharply in large-scale or high-mobility settings, whereas timing- and RSS-based methods remain robust. Notably, in high-mobility scenarios (\emph{HMMS}), IETime and TimeRSS reduce MAE by $\sim$24\% compared to lower one (\emph{HMLS}) with similar scale, confirming that WiFiSim faithfully captures complex mobility effects. Overall, these results validate WiFiSim’s practicality for evaluating de-randomization and device counting approaches.

\section{Conclusion}
We presented \emph{WiFiSim}, a simulation framework for WiFi PR generation that integrates protocol analysis of AOSP with FSM-based device behavior modeling. Experiments showed that WiFiSim can offer complete ground truth, high fidelity, diverse behaviors, and scalable parallel simulation. In the future, we plan to extend WiFiSim with richer mobility models and broader device heterogeneity, further advancing WiFi sensing and mobility analytics.

\newpage

\section{Acknowledgments}
This work was supported in part by the National Natural Science Foundation of China under Grants 62402249, 42161070, and 62262046, in part by the Natural Science Foundation of Inner Mongolia A.R. of China under Grant 2023MS06004, in part by the Program for Young Talents of Science and Technology in Universities of Inner Mongolia A.R. of China under Grant NJYT25011, in part by the Ordos Science \& Technology Plan Grant YF20240029, in part by the fund of supporting the reform and development of local universities (Disciplinary construction), and in part by the fund of First-class Discipline Special Research Project of Inner Mongolia A.R. of China under Grant YLXKZX-ND-036.

\bibliographystyle{IEEEbib}
\bibliography{refs}

\end{document}